\documentclass[aps,prl,twocolumn,groupedaddress,showpacs]{revtex4-1}
\usepackage{graphicx}
\usepackage{amsmath}
\usepackage{amssymb}

\begin{document}
\title{Weak and Strong Coupling Theories for Polarizable Colloids and Nano-Particles}

\author{Amin Bakhshandeh}
\affiliation{Instituto de F\'isica, Universidade Federal do Rio Grande do Sul, Caixa Postal 15051, CEP 91501-970, Porto Alegre, RS, Brazil}
\affiliation{Department of Physical Chemistry, School of Chemistry, University College of Science, University of  Tehran, Tehran 14155, Iran}

\author{Alexandre P. dos Santos }
\affiliation{Instituto de F\'isica, Universidade Federal do Rio Grande do Sul, Caixa Postal 15051, CEP 91501-970, Porto Alegre, RS, Brazil}

\author{Yan Levin}
\email{levin@if.ufrgs.br}
\affiliation{Instituto de F\'isica, Universidade Federal do Rio Grande do Sul, Caixa Postal 15051, CEP 91501-970, Porto Alegre, RS, Brazil}

\begin{abstract}
A theory is presented which allows us to accurately calculate the density profile of monovalent and multivalent counterions in  suspensions of polarizable colloids or nano-particles.  In the case of  
monovalent ions, we derive a weak-coupling theory that explicitly accounts for the ion-image interaction, leading to a modified
Poisson-Boltzmann equation.  
For suspensions with multivalent counterions, a strong-coupling theory is used
to calculate the density profile near the colloidal surface and a Poisson-Boltzmann equation with a renormalized boundary condition to account for the counterion distribution in the far-field. All the results are compared with the 
Monte Carlo simulations, showing an excellent agreement between the theory and the simulations. 
\end{abstract}


\maketitle

Colloidal suspensions are of great practical interest for biology, chemistry, and physics. 
The subject has a long history going back more than a hundred years. In spite of the intense effort, many interesting phenomena
which are found in colloidal science have not been fully elucidated.  
For example, it is well known that the stability of a hydrophobic colloidal suspension depends specifically on 
the electrolyte present in suspension.   Addition of multivalent counterions results in a rapid precipitation
of colloidal particles.  What is more surprising is that even for monovalent counterions, 
stability of colloidal suspensions depend strongly on the precise nature of the counterions.  Thus, as one goes along the halogen
series, the critical coagulation concentrations of positive colloidal particles can decrease by as much as an order of magnitude,
when anion is changed from fluoride to iodide~\cite{DoLe11}.  Another interesting phenomenon found in suspensions with multivalent ions 
is the reversal of electrophoretic mobility~\cite{QuCa02,FeFe05}, or equivalently charge reversal~\cite{LoSa82,PiBa05,GuGo10,DoDi10}. Under some conditions, it is  also possible to observe like-charge 
attraction between the colloidal particles 
of the same sign of charge~\cite{RoBl96,LiLo99,WuBr99,GeBr00,SoDe01}.  Many of these interesting phenomena are the consequence of strong
electrostatic correlations between the counterions.  
The role of electrostatic correlations has been studied using simple models of colloidal suspensions which neglect
particle polarizability.
The standard Poisson-Boltzmann equation (PB) ---  used extensively in colloidal science ---
fails to account for the induced charge at the particle-solvent interface,
predicting that the counterion density should remain unaffected by the colloidal polarizability.
In this Letter,  we will show that the induced colloidal charge significantly modifies the ionic density distribution
even for monovalent counterions. The theory developed in this Letter allows us to accurately predict the counterion density
distribution both in the weak (monovalent counterions) and strong (multivalent counterions) coupling limits.

We will use the primitive model of colloidal suspension in which colloidal particles are represented by hard spheres
of radius  $a$ and dielectric constant $\epsilon_c$ with the charge  $-Zq$ distributed uniformly over the surface.  
Water is  modeled as a uniform dielectric of 
permittivity $\epsilon_w$. The system is at room temperature, so that the Bjerrum length, 
defined as $\lambda_B=q^2/\epsilon_w k_BT$, is $7.2$\AA.  To account for the finite concentration of colloidal particles,
we will use a spherical Wigner-Seitz~(WS) cell of radius $R$. 
The cell also contains  $N=Z/\alpha$ $\alpha$-valent counterions, each of radius $r_c$.  For most colloidal suspensions of 
practical interest $\epsilon_w/\epsilon_c \gg 1$.  In this limit it is possible to show that the exact Green function for the 
interaction between two counterions is~\cite{Dos},
\begin{equation}
G({\bf r},{\bf r'})=\frac{1}{\epsilon_w |{\bf r}-{\bf r'}|}+\frac{ a}{\epsilon_w r'|{\bf r}-\frac{a^2}{r'^2}{\bf r'}|}+\psi_c({\bf r},{\bf r'})  \ .
\label{1}
\end{equation}
The terms in Eq.~\ref{1} are respectively the electrostatic potential produced by an ion located at ${\bf r'}$, image charge located at the inversion point $\frac{a^2}{r'^2}{\bf r'}$ inside the colloid, 
and the counter-image charge spread uniformly from the inversion point 
up to the center of the colloidal particle~\cite{Norris,Lindell},
\begin{equation}
\psi_c({\bf r},{\bf r'})=\dfrac{1}{\epsilon_w a}
\log \left(\frac{r r' -{\bf r} \cdot {\bf r'}}{a^2-{\bf r} \cdot {\bf r'}+\sqrt{a^4-2 a^2 ({\bf r} \cdot {\bf r'})+r^2 r'^2}}\right) \ .
\label{app}
\end{equation}
Eq.~\ref{1} can be used to obtain the counterion density distributions using Monte Carlo~(MC) as was shown in Ref. \cite{Dos}, and is a much faster alternative to simulations based on expansion in Legendre polynomials~\cite{Me02}.

To calculate the density profiles theoretically one often relies on the mean-field PB equation.  In the case of non-polarizable
colloidal particles, it is well known that for monovalent counterions  the PB theory is very accurate~\cite{Levin}.  On the other hand,
for multivalent counterions there are strong deviations from the prediction of PB equation~\cite{Levin,shklovskii,NaJu05,DoDi09,samaj,KaDo09}.
These deviations
are a consequence of strong electrostatic correlation between the multivalent 
counterions close to the colloidal surface. For polarizable particles, however, the usual PB equation fails even for
monovalent counterions.  In this case, the failure of the PB equation is a consequence of the counterion-image interaction --- i.e.
transverse correlations --- which are left out of the mean-field PB equation.  The deviations
from PB theory augment with increasing counterion valence.  One of the objectives of the present work 
is to construct a theory that can properly take into account both the ion-ion and the ion-image interactions for polarizable colloidal particles.  

The electrostatic potential inside the WS cell satisfies the exact Poisson equation
\begin{equation}
\nabla^2\phi(r)=\frac{Zq}{\epsilon_w a^2}\delta(r-a)-\frac{4\pi}{\epsilon_w}\alpha q\rho(r) \,,
\label{pb}
\end{equation}
where  $\phi(r)$ is the mean electrostatic potential at distance $r$ from the colloidal center and $\rho(r)$ is the
mean counterion density.  We next suppose that the counterions are distributed in accordance with the Boltzmann distribution
\begin{equation}
\rho(r)=\frac{Z e^{-\beta\alpha q\phi(r)-\beta W(\kappa,z)}}{4 \pi \alpha \int_{a+r_c}^R dr \ r^2 e^{-\beta\alpha q\phi(r)-\beta W(\kappa,z)}} \, ,
\label{dist}
\end{equation}
where $\beta=1/k_B T$ and $z$ is the distance from the colloidal surface, $z=r-a$. In writing Eq.~\ref{dist}, we have implicitly assumed that the main contribution to the potential of mean-force, besides the electrostatic potential $\phi(r)$, comes from 
the ion-image interaction, $W(\kappa,z)$.  Although this is true for the monovalent counterions --- weak-coupling limit --- 
this assumption breaks-down for the multivalent ions in the strong-coupling limit.  

Unfortunately there is no exact way of calculating the
charge-image interaction, so that approximations must be invoked.  We proceed as follows.  
Consider a neutral one component plasma (OCP) confined to a half-space by a hard
dielectric wall of $\epsilon_c \approx 0$.  For $z<r_c$ the hard core repulsion requires that the electrostatic potential satisfies the Laplace equation $\nabla^2\phi(r)=0$, where we have assumed that
the neutralizing background of the OCP also starts at $z=r_c$.   For $z>r_c$ the electrostatic potential inside the OCP satisfies the linearized PB equation $\nabla^2\phi(r)=\kappa^2 \phi(r)$, where $\kappa=\sqrt{\frac{4 \pi \alpha \lambda_B Z}{ V}}$ and $V$ is the volume accessible to the counterions. These two equations can be solved explicitly
to calculate the work that must be done to bring an ion of the OCP from infinity to the distance $z$ 
from the wall~\cite{Levin1}.  We obtain 
\begin{equation}
\beta W_{0}(\kappa,z) \approx \frac{\beta W_0(\kappa,r_c)r_c}{z}e^{-2\kappa(z-r_c)} \, ,
\label{wap}
\end{equation}
where 
\begin{equation}
\beta W_0(\kappa,r_c)=\frac{\alpha^2 \lambda_B }{2}\int_0^\infty dk\frac {k[s\ \mathrm{cosh}(k r_c)-k\ \mathrm{sinh}(k r_c)]}{s[s\ \mathrm{cosh}(k r_c)+k\ \mathrm{sinh}(k r_c)]} 
\label{4}
\end{equation}
and $s=\sqrt{k^2+\kappa^2}$.  Eq.~\ref{wap} accounts for two fundamental contributions:  the ion-image interaction which
is screened with the Debye length $1/\kappa$, and for the loss of the electrostatic solvation free energy experienced by the ion near
the interface.  Since the wall has the dielectric constant much smaller than that of water, the ion-image interaction is
strongly repulsive.  However, note that even if there would not be any dielectric discontinuity but simply a hard wall, 
there still would be an electrostatic 
energy penalty for bringing an ion from the bulk of electrolyte --- where its self-energy is efficiently screened by the background and
by the other ions --- to the interface, where the broken symmetry prevents the efficient screening of its electric field, 
resulting in a higher electrostatic self-energy. 
Therefore, besides the ion-image repulsion, at finite ionic concentrations there is 
an additional repulsive interaction arising from the broken translational symmetry near the wall-water interface, which is also
taken into account in Eq.~\ref{wap}.  

\begin{figure}[h]
\begin{center}
\includegraphics[width=6.5cm]{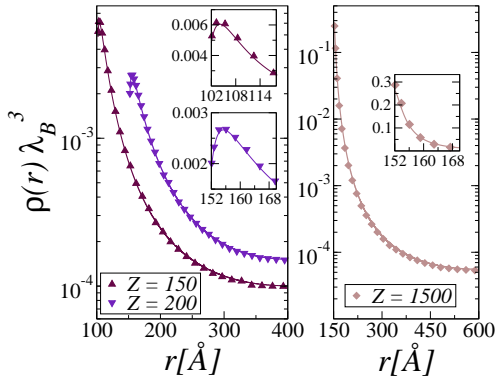}
\end{center}
\caption{Density profiles of monovalent counterions. Symbols represent simulation data and lines represent the modified PB theory. The parameters of the simulations are $\alpha=1$, $r_c=2$\AA; down triangles ($a=150$\AA\ and $R=400$\AA); up triangles ($a=100$\AA\ and $R=400$\AA); diamonds ($a=150$\AA\ and $R=600$\AA). The insets show the profiles close to the colloidal surface.}
\label{fig1}
\end{figure}

It is instructive to rewrite Eq.~\ref{wap} as
\begin{equation}
\beta W_0(\kappa,z)=\left[\Delta_{sc}(\kappa,z)+ \frac{\beta W_0(0,r_c)r_c}{z}\right]e^{-2\kappa(z-r_c)} \ .
\label{wap2}
\end{equation}
The first term in the square brackets, $\Delta_{sc}(\kappa,z)=\beta W_0(\kappa,r_c)r_c/z-\beta W_0(0,r_c)r_c/z$ with
$\beta W_0(0,r_c)=\alpha^2 \lambda_B/4 r_c$,
is the repulsive energy arising from the 
broken translational symmetry, while the second term is due to the ion-image interaction.  For colloids and
nano-particles, we expect that $\Delta_{sc}(\kappa,z)$ depends only weakly on the curvature and can be approximated by
that of a planar wall. 
On the other hand, the ion-image interaction energy is strongly dependent on the radius of curvature of the dielectric interface
and must be corrected when treating nano-particles or colloids. Using the Green function Eq.~\ref{1},
the curvature correction can be easily calculated.  We find
\begin{equation}
\beta W(\kappa,z)=\left[\Delta_{sc}(\kappa,z)+\frac{\alpha^2 a \lambda_B}{2 (z^2+2az)}+\frac{\beta \psi(z)}{2}\right]e^{-2\kappa(z-r_c)} \ ,
\label{wapc}
\end{equation}
where
\begin{equation}
\beta \psi(z)=\dfrac{\alpha^2 \lambda_B}{a} \log \left[ 1-\frac{a^2}{(a+z)^2} \right] \ .
\label{wself}
\end{equation}
These expressions are used in Eqs.~\ref{pb} and \ref{dist} to account for the colloidal polarizability.
The modified PB equation is then solved numerically to calculate the counterion density profiles.
In Fig.~\ref{fig1} the theory is compared with the MC simulations for monovalent ions, showing an excellent agreement. 
On the other hand, in the Supplementary Information, we show that the usual PB equation  deviates significantly
from the MC data near the colloidal surface.

Although we find a perfect agreement between the theory and the simulations for monovalent counterions, 
strong deviations are observed when multivalent ions are present in suspension~\cite{pincus}.
This is similar to what has been previously found with
non-polarizable colloidal particles~\cite{DoDi09}.
Near the colloidal surface, ionic concentration is very large and the electrostatic interaction between the 
multivalent counterions is very strong.  This results in a 
formation of a strongly correlated quasi-two-dimensional OCP on the surface of colloidal particle. 
The strength of electrostatic correlations can be quantified by the plasma parameter  
$\Gamma=\frac{\alpha^{3/2}\lambda_B \sqrt{Z}}{2(a+r_c)}$, which is the ratio of the characteristic electrostatic to the
thermal energy of condensed counterions~\cite{Levin}.  For $\Gamma \gg 1$ the counterions exhibit local hexagonal order.  

\begin{figure}[t]
\begin{center}
\includegraphics[width=8.cm]{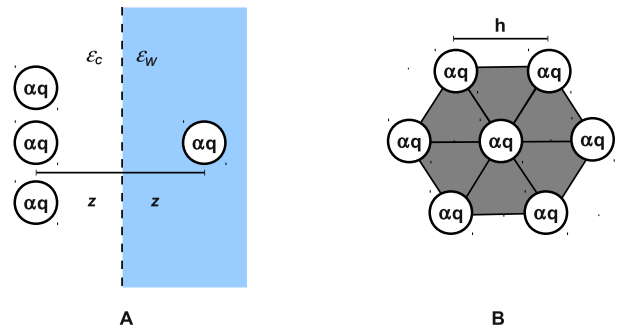}
\end{center}
\caption{Hexagon of images at the surface. In (A) the side view. In (B) the self-image and the nearest neighbors.}
\label{fig2}
\end{figure}

\begin{figure}[b]
\begin{center}
\includegraphics[width=6.5cm]{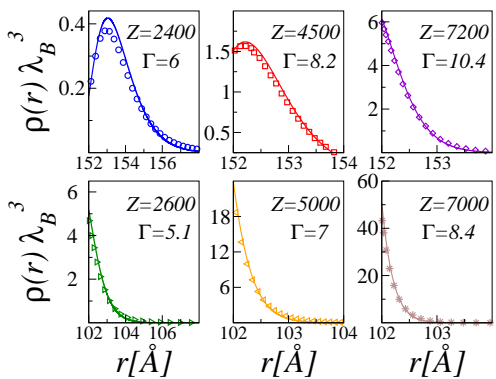}
\end{center}
\caption{Density profiles near the colloidal surface. Symbols are the simulation data and lines are the predictions of the present theory. Circles, squares and diamonds are simulation data for $\alpha=3$, $a=150$\AA\ and $R=600$\AA. Right, left triangles and stars are simulation data for $\alpha=2$, $a=100$\AA\ and $R=400$\AA. The ionic radius is $r_c=2$\AA.}
\label{fig3}
\end{figure}

Consider a counterion of the condensed layer.  If the average separation between the condensed counterions is much smaller
than the radius of the nano-particle, the effects of the curvature will be screened.  Thus, in the limit 
$\Gamma \gg 1$, $a \gg r_c$, and  $\frac{\sqrt{\pi} \alpha^2 \lambda_B}{(a+r_c) \Gamma} \ll 1$,  we can neglect the curvature of the colloidal surface and treat it as a hard
wall separating the low dielectric $\epsilon_c \approx 0$ half-space from the high-dielectric region 
occupied by the counterions, Fig.~\ref{fig2}.   
As a consequence of the 
hexagonal symmetry, the electric field produced on the counterion by other condensed counterions vanishes.  
Therefore, in the limit
$\Gamma \rightarrow \infty$,
the ion interacts only with the electric field of the charged wall and the field produced by the images. 
The average separation $h$ between the condensed counterions of the hexagonal lattice is 
$h=\sqrt{8\pi (a+r_c)^2 \alpha/(Z\sqrt{3})}$,  considering that all $Z/\alpha$  are in the area $4\pi (a+r_c)^2$. 
In the strong coupling limit, the counterions density distribution then take a particularly simple form 
\begin{equation}
\rho_{sc}(z)=A \ e^{-2 (z-r_c)/l_{gc}-\beta W_h(z) } \ ,
\label{sc}
\end{equation}
where  $l_{gc}=\frac{2a^2}{Z\alpha \lambda_B}$ is the Gouy-Chapman length and $ W_h(z)$ is the electrostatic potential
of interaction between the ion and its image and the images of its first nearest neighbors,
\begin{equation}
\beta W_h(z)=\frac{\alpha^2 \lambda_B}{ 4 z} + \frac{6\alpha^2 \lambda_B}{\sqrt{(2 z)^2 + h^2}} \ .
\label{hex}
\end{equation}
The first term in the exponential of Eq. \ref{sc} is due to the ion interacting with the field of the charged hard wall.  
Note that because 
$\epsilon_c \approx 0$, the displacement field inside the low-dielectric half-space vanishes, which leads to the factor of $2$ in the exponent of Eq.~\ref{sc}, that is absent in the strong-coupling theories of non-polarizable particles~\cite{DoDi09}.
The normalization factor $A$ is obtained using the charge neutrality condition,
\begin{equation}
A=\frac{Z/\alpha}{4 \pi (a+r_c)^2 \int_{r_c}^{R-a}  dz \  e^{-2 (z-r_c)/l_{gc}-\beta W_h(z) }} \ .
\label{10}
\end{equation}
In Fig.~\ref{fig3} we show the density distributions for different colloidal systems containing either divalent or trivalent counterions, $\alpha=2$ and $\alpha=3$, respectively.
A very good agreement is observed between the theory and the simulations.  
As expected, the agreement improves for larger $\Gamma$ and when $a \rightarrow \infty$ (see Supplementary Information). 

\begin{figure}[h]
\begin{center}
\includegraphics[width=6.5cm]{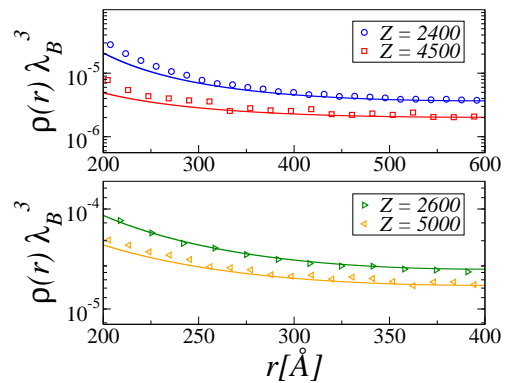}
\end{center}
\caption{Density profiles in the far-field. Symbols represent the simulation data and lines are the predictions of 
the present theory. The parameters are the same as in Fig.~\ref{fig3}.}
\label{fig4}
\end{figure}

Far from the colloidal surface the concentration of counterions drops rapidly.  In this far-field region
the counterion correlations can be neglected and the mean-field Poisson-Boltzmann theory once again becomes applicable.  
To connect the strongly-correlated region near the colloidal surface with the weakly-correlated far-field, 
we use the theory developed in Ref.~\cite{DoDi09}. The uniformity of the chemical potential throughout the system
requires that the counterion concentration at the boundary of the WS cell be related with the coarse-grained density 
in the strongly-correlated region described by Eq.~\ref{sc} resulting in equation
\begin{equation}
\rho_{bc}(R)=\rho_{cg}e^{\beta \alpha q \left[\phi(a+r_c)-\phi(R)\right] + \beta \mu_c} \ ,
\label{nbc}
\end{equation}
where $\beta \mu_c=-1.65\Gamma + 2.61\Gamma^{1/4} -0.26 \ln \Gamma -1.95$ is the chemical potential of the strongly-correlated 2d OCP~\cite{tot} and $\rho_{cg}$ is the coarse-grained density of the 2d OCP,
\begin{equation}
\rho_{cg}=\frac{\int_{r_c}^{r_c+l_{sc}} dz \ \rho_{sc}(z)}{l_{sc}} \ ,
\label{cg}
\end{equation}
where $l_{sc}=3.6\ l_{gc}$~\cite{DoDi09}.  Solving the usual PB equation while enforcing the condition Eq.~\ref{nbc} at the cell 
boundary, we obtain both the effective colloidal charge and the counterion density 
profiles in the weakly-correlated region. In Fig.~\ref{fig4}, we show the comparison between MC simulations and the predictions of the present theory in the far-field region.  Once again
an excellent agreement is found between the theory and the simulations.

We have derived a new theory which allows us to quantitatively calculate the density distributions of monovalent and multivalent counterions in suspensions of polarizable colloidal particles. In the case of the monovalent ions, we have 
derived a modified PB equation
which allows us to calculate the complete counterion density profile. 
For multivalent ions, we used the strong-coupling theory to
obtain the density profiles near the colloidal surface 
and the usual PB equation with the renormalized boundary condition to calculate the density distribution in the far
field.  All the results are in excellent agreement with the MC simulations. Finally, we mention that the same theory
can be easily extended to treat metal nano-particles.

This work was partially supported by the CNPq, Fapergs, INCT-FCx, and by the US-AFOSR under the grant FA9550-09-1-0283.
 
\bibliography{ref.bib}

\end{document}